# A Data-Driven Paradigm-Based Image Denoising and Mosaicking Approach for High-Resolution Acoustic Camera

○Xiaoteng Zhou, Yilong Zhang, Katsunori Mizuno (The University of Tokyo),

Kenichiro Tsutsumi, Hideki Sugimoto (Penta-ocean construction Co.,LTD)

*Abstract*—In this work, an approach based on a data-driven paradigm to denoise and mosaic acoustic camera images is proposed. Acoustic cameras, also known as 2D forward-looking sonar, could collect high-resolution acoustic images in dark and turbid water. However, due to the unique sensor imaging mechanism, main vision-based processing methods, like image denoising and mosaicking are still in the early stages. Due to the complex noise interference in acoustic images and the narrow field of view of acoustic cameras, it is difficult to restore the entire detection scene even if enough acoustic images are collected. Relevant research work addressing these issues focuses on the design of handcrafted operators for acoustic image processing based on prior knowledge and sensor models. However, such methods lack robustness due to noise interference and insufficient feature details on acoustic images. This study proposes an acoustic image denoising and mosaicking method based on a data-driven paradigm and conducts experimental testing using collected acoustic camera images. The results demonstrate the effectiveness of the proposal.

## I. Introduction

In recent years, to meet the demands of human sustainable development, activities such as coastal infrastructure construction, ecological investigation, and energy exploration have gradually extended to deepwater. Deepwater areas are usually characterized by low-visibility conditions, including darkness and turbidity. In low-visibility environments, traditional optical cameras cannot work well due to insufficient illumination and medium interference.

To solve the perception challenge in deepwater, high-resolution acoustic camera sensors [1] such as dual-frequency identification sonar (DIDSON) and adaptive resolution imaging sonar (ARIS) have become popular visual devices. The acoustic camera is an active forward-looking sonar (FLS) that can emit multiple 3D fan-shaped sound waves and receives reflected signals when they collide with objects in the forward direction, thereby generating high-resolution acoustic images [1].

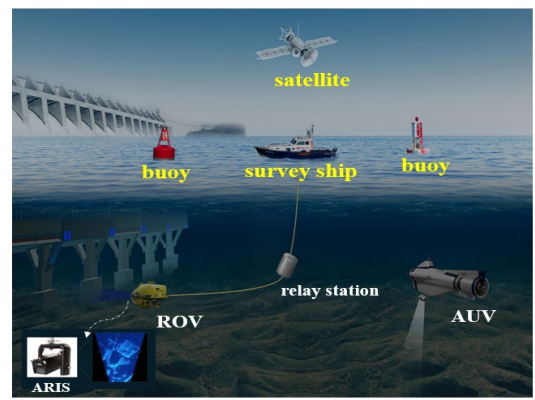

Fig. 1  Marine application demo of acoustic cameras.

Nevertheless, given the unique acoustic imaging mechanism and sensor model, acoustic cameras face two major constraints, namely noise interference and narrow field of view. Compared with improving the sensor hardware system, these issues could be solved quickly by applying basic acoustic image processing algorithms such as image denoising and mosaicking. Although these algorithms have made rapid progress in optical sensing, their application to acoustic images shows a significant lag. Taking image denoising as an example, this step is a key preprocessing step for most image-based vision applications. However, it still poses challenges when applied to acoustic images.

In past research, processing of acoustic images has focused on designing handcrafted operators using prior knowledge and sensor models [2,3]. However, these methods lack robustness and often fail due to noise and scene changes. Furthermore, due to the complex characteristics of acoustic images, designing handcrafted operators often requires acquiring prior knowledge, presenting a challenge for newcomers. With the continuous development of computer vision technology, there is now an opportunity to introduce deep learning approaches based on data-driven paradigms to process acoustic images, eliminating the need to design complex handcrafted operators.

In summary, the research in this work proposed a novel image denoising and mosaicking approach for acoustic cameras based on data-driven paradigms and validated the effectiveness through real experiments.

## II. RESEARCH METHODOLOGY

### A. Acoustic Camera Images Denoising

Image denoising is a prerequisite step for image-based vision applications because the distribution of noise will affect the performances of downstream algorithms. A robust image denoising algorithm can enhance the readability of acoustic camera images, and can also improve the overall effect of mosaicking. Since practical acoustic images are generally affected by noise, it is quite difficult to prepare noise-clean image pairs for the training of supervised denoising models. Therefore, this study adopts a self-supervised training strategy [4] that relies only on noisy images to train denoising models, as shown in Fig. 2.

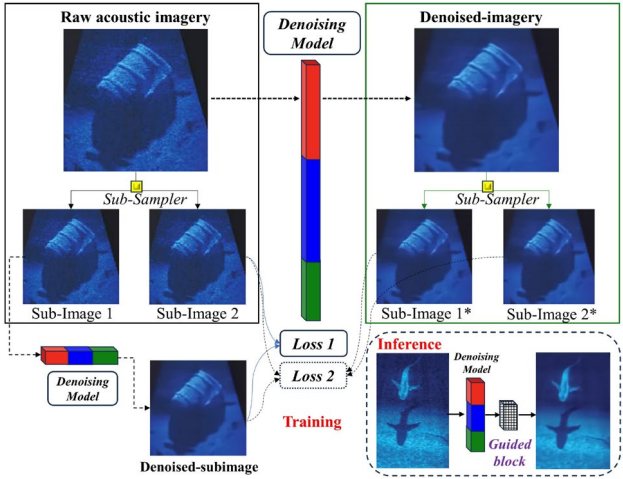

Fig. 2 Pipeline of self-supervised denoising strategy.

In this study, an acoustic image training dataset was made using image data from the Sound Metrics [5]. The training of the self-supervised denoising model was achieved by acquiring noise image pairs through the sub-pixel sampling of the noisy images in the dataset. The strategy proposed is applicable to images of any shape, with acoustic camera images often presented in a wedge-shaped format. In addition, it requires fewer training samples, making it suitable for acoustic image processing that lacks enough datasets.

Fig. 2 illustrates the training pipeline for self-supervised denoising of acoustic images. The body section is the complete view of the training. Generate a pair of images from noisy acoustic images using adjacent subsamplers. The denoising network model uses the subsampled images as input and target. The loss consists of two parts: the upper part calculates the $Loss1$ between the network output and the noise target; the lower part, the $Loss2$ is further added considering the difference between the subsampled noisy image and the ground-truth value [4]. It should be mentioned that neighborhood subsamplers (block in yellow) appearing twice are the same. The bottom right part is an inference demo using the trained acoustic denoising model. The loss function for network training is defined as follows:

$$Loss = Loss1 + \gamma \cdot Loss2 \quad (1)$$

$$Loss1 = \|Denoise(SubImage1) - SubImage2\|_2^2 \quad (2)$$

$$Sub* = (SubImage1* - SubImage2*) \quad (3)$$

$$Loss2 = \|Denoise(SubImage1) - SubImage2 - Sub*\|_2^2 \quad (4)$$

where $Denoise(\theta)$ is the denoising function, for a raw noisy acoustic image, two sub-samples $SubImage1$ and $SubImage2$ are taken. The $Loss1$ is computed between the denoised image $Denoise(SubImage1)$ obtained using the trained denoising model and $SubImage2$, which represents the target noise. It could calculate the reconstruction error between the network output and the noise target. And, a regularization term is introduced in $Loss2$ to alleviate the over-smoothing of the output image caused by the sampling method. $\gamma$ is a hyperparameter that controls the strength.

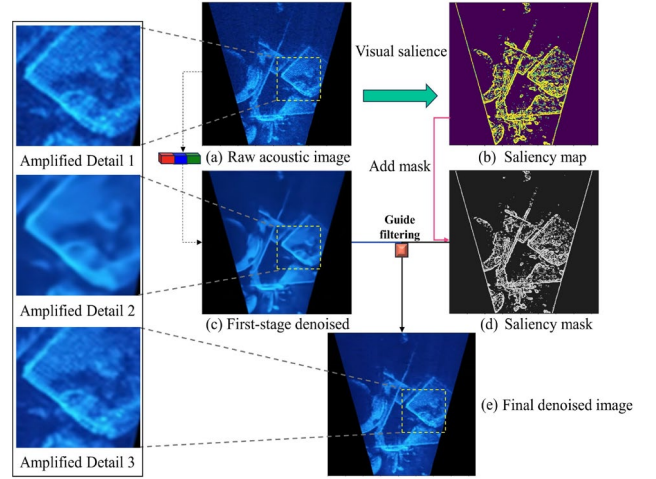

Fig. 3 Feature-guided block applied on acoustic images.

To better preserve key feature details on acoustic images while filtering noise, this study designed a fine feature-guided block to enhance the denoising model output (first-stage denoised) through guided filtering [6] and visual saliency detection. As shown in Fig. 3, assuming $p$ represents the input noisy image, which is the raw acoustic image without any processing, an $I$ represents the guide image, which is the first-stage denoised image obtained through the self-supervised denoising strategy. Then, the feature-guided block is used to guide the transformation of the first-stage denoised image to obtain the final denoised image $q$. At this point, $q$ is nearly free from noise interference while recovering structural features, such as edges and corners. In this process, the filtering processing for one pixel of an image can be expressed as follows:

$$q_i = \sum_j W_{ij}(I) p_j \qquad (5)$$

where $q_i$ denotes the denoised pixel value at position $i$, $p_j$ represents the pixel value of the raw acoustic camera image at the position $j$, and $W_{ij}$ denotes the weight assigned to the pixel value $p_j$ based on its mapping relationship with the guide image $I$.

### B. Acoustic Camera Images Mosaicking

As shown in Fig. 4, the initialization block first parses the raw data file from the acoustic camera, transforming it into a sequence of acoustic images for subsequent processing. In the second stage, the self-supervised denoising strategy proposed is introduced to filter raw images. The third-stage mosaicking block consists of two parts: feature matching and image mosaicking. This study chooses a mosaicking pattern based on local features, mainly considering speed and robustness. A matching pattern that does not rely on a local feature detector is introduced [7] and optimized around acoustic images. This pattern eliminates the dependence on enough repeatable feature points, enabling create global matching, even in weak-texture areas. Next, spatial transformation relationships and inliers computed through feature matching results are used to solve the transformation trajectory between adjacent sonar images, and the solution results serve as input of the mosaicking pipeline. The mosaicking pipeline is divided into two steps: image fusing and mosaicking. Fusing is used to correct the stitching gaps between adjacent images, while mosaicking is used to generate high-resolution acoustic panoramas.

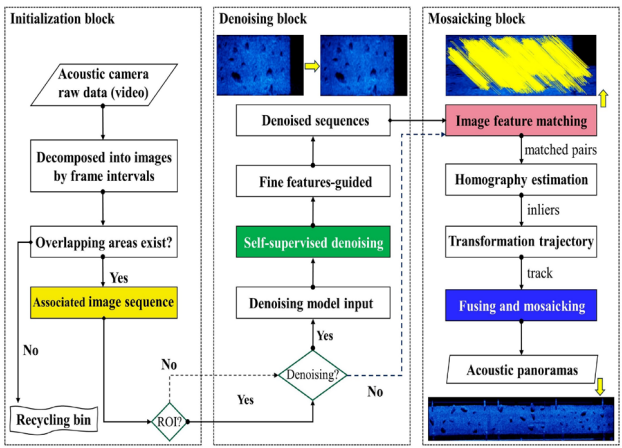

Fig. 4 Mosaicking framework that generates the large panoramas from acoustic camera imageries.

It is worth noting that the mosaicking approach proposed in this study relies solely on acoustic camera imageries without the need for additional sensor data. The input and output of the framework are as follows:

**Input**:
1) Acoustic imageries with overlapping regions.

**Output**:
1) Denoised acoustic camera imageries.
2) Acoustic camera sensor motion trajectories.
3) Large-area and high-quality acoustic panoramas.

### III. EXPERIMENT AND VERIFICATION

To validate the proposed acoustic camera image denoising and mosaicking approach in this work, acoustic images collected from on-site experiments [8] were used for testing. The experimental scenario is specifically described as follows: a concrete plate was inspected using a detection platform equipped with an ARIS Explorer 1800 acoustic camera at one detection distance of 1.5 meters, as depicted in Fig. 5(a) to (c). Fig. 5(d) and (e) showcase the raw data files and model of the acoustic camera. The images obtained after preprocessing are shown in Fig. 5(f) and (g).

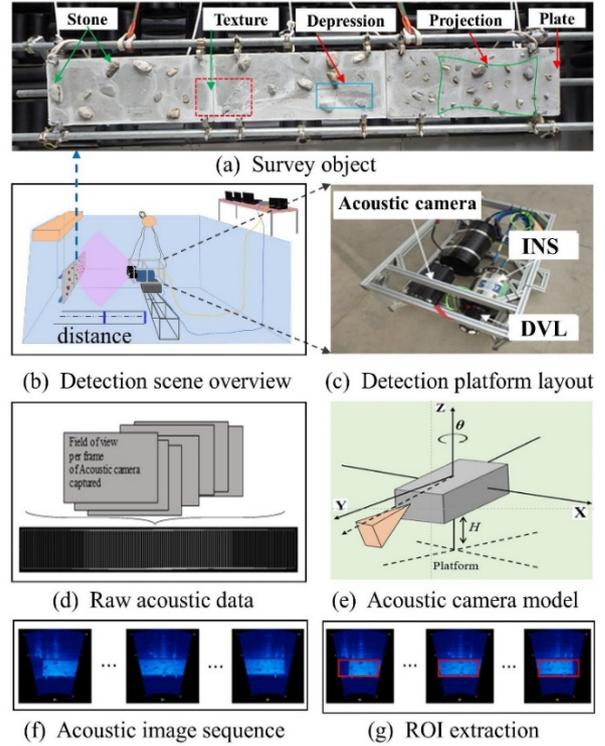

Fig. 5 Experimental site for acoustic image collection.

The collected acoustic camera images are first used as a testing dataset for the denoising approach, and then input into the mosaicking framework to generate acoustic panorama. Firstly, this section compares the performance of the proposed denoising method with common denoising methods, including classic filters and deep learning-based denoising models, such as NBR2NBR model [4], as shown in Fig. 6. Due to space limitations, this section randomly selects one image frame for display. It can be seen that our approach could effectively filter out the noise on the acoustic image while better preserving the feature details.

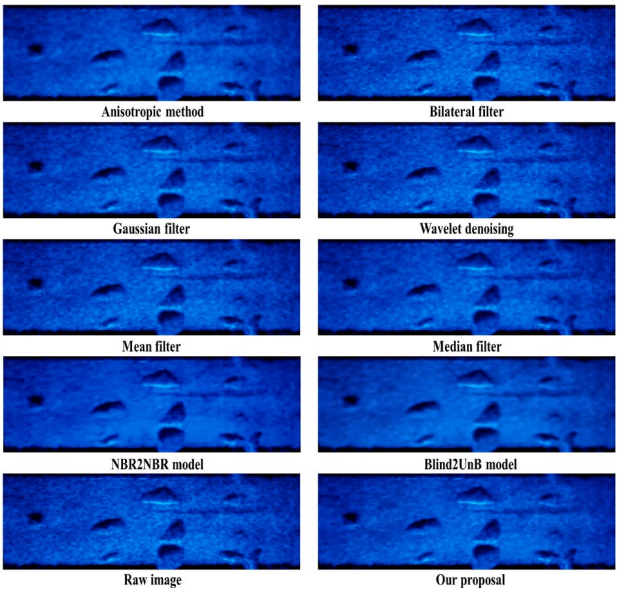

Fig. 6 Comparison of the denoising performances.

Subsequently, feature matching is performed on the denoised acoustic images and the matching results are used for mosaicking, as shown in Fig. 7. The process of image feature matching is like stitching the consistent features on adjacent frames with a needle and thread to generate a complete panorama. In this process, the positions of the feature points act as the pinhole, and the matching results serve as the threads. Many studies believed that the handcrafted feature-matching operators are not suitable for acoustic images due to noise and weak texture features [3]. Thus, this work introduced a novel feature-matching pattern based on a data-driven paradigm [7] to achieve the robust matching of acoustic camera images.

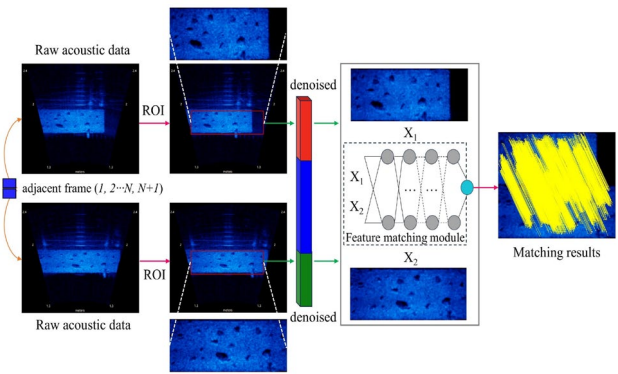

Fig. 7 Feature matching between acoustic image pairs.

When the image feature matching results between adjacent frames are obtained, the geometric mapping relationship can be calculated by the module, and then the accurate mosaicking results can be generated.

In the end, the proposed mosaicking approach was used to recover the concrete plate detected, as shown in Fig. 8. It can be seen that our proposal can robustly restore the concrete plate appearance and highlight features (e.g. particles and depressions), and does not require any external sensor data.

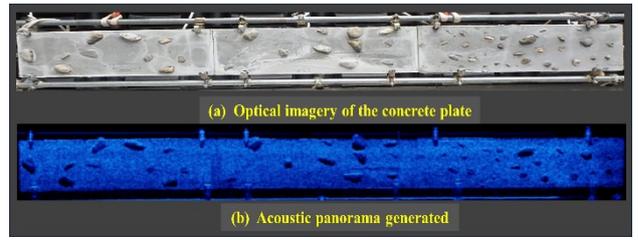

Fig. 8 Mosaicking results of acoustic image sequences.

## IV. CONCLUSIONS

This work proposes a denoising and mosaicking approach based on data-driven paradigms specifically applied to acoustic camera images, which is validated on practical experiment. The results show that the denoising approach can effectively filter out noise while preserving the fine feature details of acoustic images. It does not need to make a priori assumptions on the noise model and does not need complex post-processing. The mosaicking method could overcome the challenge of sparse features on acoustic images and is able to robustly generate acoustic panoramas.

In future work, we will investigate the impact of target materials on the image denoising algorithm, as materials directly affect acoustic imaging. We will also explore the impact of detection distance and sensor posture on the mosaicking performance. This work aims to consolidate the fundamental acoustic image processing, and then provide support for expanding the real application of acoustic cameras.